\documentclass[prb,preprint]{revtex4-1} 

\usepackage{amsmath} 
\usepackage{amsfonts} 
\usepackage{graphicx} 
\usepackage{hyperref}
\usepackage{cleveref}
\usepackage{float}
\begin{document}

\title{Coherence and path indistinguishability for the interference of multiple single-mode fields }

\author{Rathindra Nath Das}
\email{rathin.phy@iitb.ac.in} 
\affiliation{Department of Physics, Indian Institute Technology Bombay, Powai, Mumbai, Maharashtra 400076}

\author{Sobhan Kumar Sounda}
\email{sobhan.physics@presiuniv.ac.in}
\affiliation{Department of Physics, Presidency University, 86, 1, College St, College Square, Kolkata, West Bengal 700073}

\date{\today}

\begin{abstract}
A well known result for the interference of two single-mode fields is that the degree of coherence and the degree of indistinguishablity are same when we consider the detection of a single photon. In this article we present the relation between degree of coherence, path indistinguishability and the fringe visibility considering interference of  multiple number of single-mode fields while being interested in the detection of a single photon only . We will also mention how Born's rule of interference for multiple sources is reflected in these results.
\end{abstract}

\maketitle

\section{Introduction} 

The equivalence of coherence and indistinguishability is an very important result in the perspective of wave-particle duality. Coherence is one of the main criteria for interference of light beams. On the other hand in single photon interference experiment the lack of photon's path information plays a crucial role. In his famous article L. Mandel\cite{Mandel} has shown that the modulus of degree of second order coherence is identical to the degree of indistinguishability for the case of two single-mode fields emitted from two sources and by considering the detection a single photon only. In recent times experiments of interference for more than two slits\cite{Triple} has been of prime interest for proving the validity of Born's rule . In this paper we describe the generalisation of L. Mandel's result for 3 single-mode fields followed by N number of single-mode fields while considering the detection of a single photon only. We also discuss how the equivalent form of Born's rule is obtained in the context of coherence and fringe visibility.

\section{Interference of three single-mode fields}
\begin{figure}[H]
\centering
\includegraphics[width=4in]{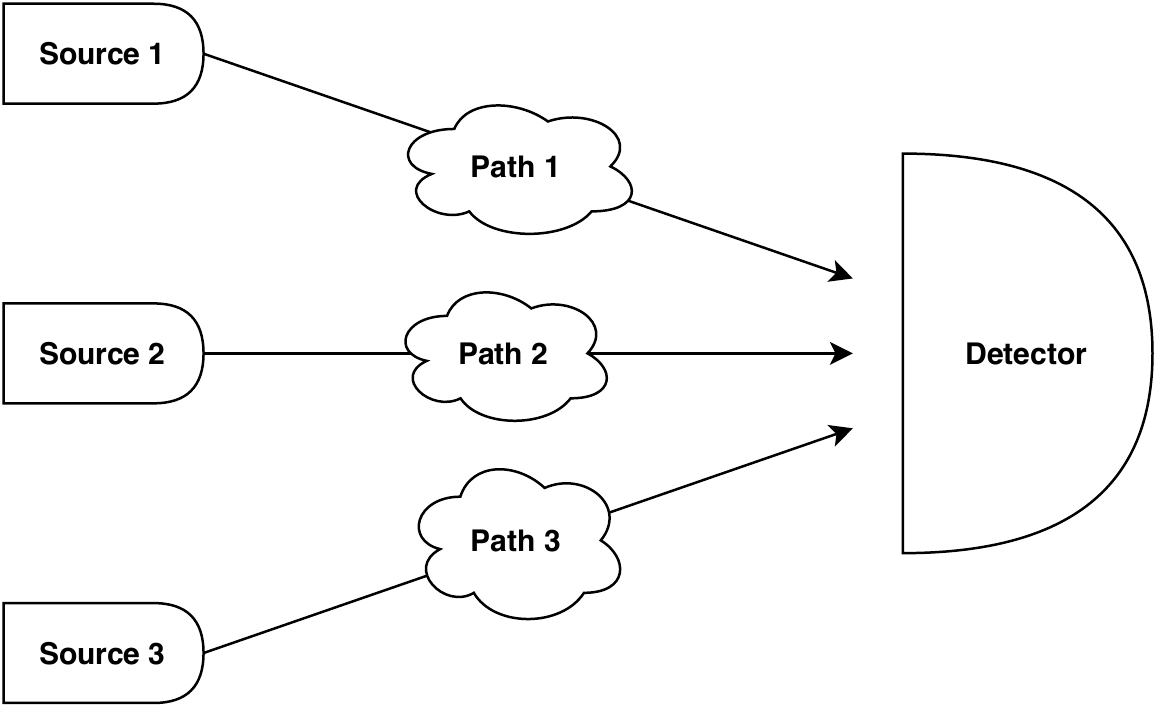}
\caption{Schematic diagram of the interference experiment}
\label{fig1}
\end{figure}

Before we discuss the generalised version of the interference, We consider the case of interference of three single-mode fields and the detection of a single photon only. The schematic diagram of the experiment under consideration is shown at Fig.\ref{fig1}. Here we have three sources and so a photon detected at any point on the detector may take three possible paths.
We write the state of the photon before the detection process takes place as \begin{align}
    |\psi \rangle = \alpha |1\rangle_1|0\rangle_2|0\rangle_3 + \beta |0\rangle_1|1\rangle_2|0\rangle_3 + \gamma |0\rangle_1|0\rangle_2|1\rangle_3
\end{align}
where $|\alpha|^2$, $|\beta|^2$ and $|\gamma|^2$ are the probability of the photon being produced by the first, second and third source respectively and $|\alpha|^2+|\beta|^2+|\gamma|^2=1$. Considering this state the density matrix of this pure state can be written as \begin{align}
    \hat{\rho}_{\text{ID}} &= |\psi\rangle\langle\psi| \nonumber\\ 
    &= \begin{pmatrix} 
    |\alpha|^2 & \alpha\beta^* & \alpha\gamma^* \\
    \alpha^*\beta & |\beta|^2 & \beta\gamma^* \\
    \alpha^*\gamma & \beta^*\gamma & |\gamma|^2
    \end{pmatrix}
\end{align}
 On the other hand we could have an incoherent mixture of states and the corresponding density matrix of this photon represented by a diagonal density matrix of the form \begin{equation}
      \hat{\rho}_{\text{D}}  = \begin{pmatrix} 
    |\alpha|^2 & 0 & 0 \\
    0 & |\beta|^2 & 0 \\
    0 & 0 & |\gamma|^2
    \end{pmatrix}\label{6666}
 \end{equation}
In the second case we do not have the off diagonal terms implying that the intrinsic indistinguishability of the photon path is lost. So, now for the density matrix $\hat{\rho}_D$ in principle we will be able to detect the source of this photon experimentally and the interference pattern will be lost. Following the notation of L. Mandel here also the subscript of $\hat{\rho}_{ID}$ and $\hat{\rho}_D$ signifies the path indistinguishability and this potential path distinguishability for those two density matrices respectively. Now in this Hilbert space we take a general density matrix of the form \begin{equation}
 \hat{\rho} =\sum\limits_{n,m=1}^3\rho_{nm}|n\rangle\langle m| \label{1111}
 \end{equation}
and we can decompose it in the terms of $\hat{\rho}_{ID}$ and $\hat{\rho}_D$ to determine the degree of indistinguishability for the system . We write the general density matrix as  \begin{align}
   &   \hat{\rho} = P_{ID} \hat{\rho}_{ID} + P_D \hat{\rho}_{D} ~~~~~\text{where}~~~P_{ID}+P_D=1 \label{dec1}
 \end{align}
From Eq.\ref{dec1}, Eq.\ref{1111} and Eq.\ref{6666} we get \begin{align}
  \rho_{11} &= |\alpha|^2,~~\rho_{22} = |\beta|^2,~~\rho_{33} = |\gamma|^2\\  
  \rho_{12} &= P_{ID}\alpha\beta^*,~~\rho_{13} = P_{ID}\alpha\gamma^*,~~\rho_{23} = \beta\gamma^*.
 \end{align}
due to the hermiticity of the density matrix we can evaluate the lower triangular elements as the complex conjugate of the upper triangular ones. Now using very simple calculation we see that  
 \begin{align}
     P_{ID} &= \frac{\rho_{12}}{\sqrt{\rho_{11}\rho_{22}}}e^{-i\cdot \text{arg}(\rho_{12})}  = \frac{\rho_{13}}{\sqrt{\rho_{11}\rho_{33}}}e^{-i\cdot \text{arg}(\rho_{13})}  = \frac{\rho_{23}}{\sqrt{\rho_{22}\rho_{33}}}e^{-i\cdot \text{arg}(\rho_{23})} \label{PID3}\\ \nonumber &=\frac{|\rho_{21}|}{\sqrt{\rho_{11}\rho_{22}}} = \frac{|\rho_{31}|}{\sqrt{\rho_{11}\rho_{33}}} = \frac{|\rho_{32}|}{\sqrt{\rho_{22}\rho_{33}}}
 \end{align}
which indicates that $P_{ID}$ can be written in three equivalent forms where we only have pairwise terms out of the three slits.

Now we calculate the degree of coherence for the system and for that We denote the positive frequency component of the single-mode electric field of the source as \begin{equation}
       \hat{E}^{(+)}(r_j) =K \hat{a}_j~~~~~~~~~\text{where j = 1, 2, 3}
  \end{equation}. The normalised second order coherence function of the form\cite{Notation1,Notation2} \begin{align} \label{gdef}   g^{(1)}(x_i,x_j) &= \frac{G^{(1)}(x_i,x_j)}{\sqrt{G^{(1)}(x_i,x_i)G^{(1)}(x_j,x_j)}}\end{align} where \begin{equation}G^{(1)}(x_i,x_j)=\text{Tr}\{ \hat{\rho}  \hat{E}^{(-)}(x_i) \hat{E}^{(+)}(x_j)\} \end{equation} and $G^{(1)}(x_i,x_j)$ is the general second order coherence function. For the first pair of sources $(x_1,x_2)$ we get $G^{(1)}(x_1,x_2)$ of the form 
   \begin{align}  G^{(1)}(x_1,x_2) &= Tr\{ \hat{\rho}  \hat{E}^{(-)}(x_1) \hat{E}^{(+)}(x_2)\} \nonumber \\&=|K|^2\rho_{21} \label{12}\end{align}
   similarly we get, \begin{align} G^{(1)}(x_1,x_1) &=|K|^2\rho_{11} \label{11}\\G^{(1)}(x_2,x_2)&=|K|^2\rho_{22}\label{22}\end{align}
 Now from Eq.\ref{gdef},\ref{12},\ref{11} and \ref{22} we get, \begin{align}
    g^{(1)}(x_1,x_2) =  \frac{\rho_{21}}{(\rho_{11}\rho_{22})^{\frac{1}{2}}} \label{g12}  \end{align}
 We can do similar calculation for other pair of sources and we will get  \begin{align}
      g^{(1)}(x_1,x_3) &= \frac{\rho_{31}}{(\rho_{11}\rho_{33})^{\frac{1}{2}}} \label{g13}\\ g^{(1)}(x_2,x_3) &= \frac{\rho_{32}}{(\rho_{22}\rho_{33})^{\frac{1}{2}}}  \label{g23}\end{align}
 Now from Eq.\ref{PID3}, Eq.\ref{g12}, Eq.\ref{g13} and Eq,\ref{g23} we see that \begin{align}
     P_{ID} = |g^{(1)}(x_1,x_2)|, ~~P_{ID} = |g^{(1)}(x_1,x_3)| ~~\text{and}~~P_{ID} = |g^{(1)}(x_2,x_3)| \label{3pidd}
 \end{align}
 So, we see that the degree of indistinguishability is equal to the degree of coherence even for the case of interference with three sources but in a pair wise manner for all possible combinations of two sources and we also note that the degree of coherence for all pairs of sources are equal to each other.

  Now it is very straight forward to show that only the second order normalised coherence of this system is non zero. Any higher order coherence of the following form will be zero for this system;
 \begin{align}
 \nonumber    g^{(2)}(x_i,x_j;x_j,x_i) &= \frac{G^{(2)}(x_i,x_j;x_j,x_i)}{\sqrt{G^{(1)}(x_i,x_i)G^{(1)}(x_j,x_j)}}=0\end{align}
 where $G^{(2)}(x_i,x_j)=\text{Tr}\{  \hat{\rho}  \hat{E}^{(-)}(x_i) \hat{E}^{(-)}(x_j) \hat{E}^{(+)}(x_j) \hat{E}^{(+)}(x_i)\}$ is the general fourth order coherence function\cite{Notation2} or the three point fourth order coherence function
  \begin{align}
 \nonumber    g^{(3)}(x_i,x_j,x_k;x_k,x_j,x_i) &= \frac{G^{(3)}(x_i,x_j,x_k;x_k,x_j,x_i)}{\sqrt{G^{(1)}(x_i,x_i)G^{(1)}(x_j,x_j)G^{(1)}(x_k,x_k)}}=0.\end{align} So, we see that when we are interested in the detection of a single photon the modulus of degree of indistinguishability and the degree of second order two point coherence will be equal for all possible pairs of the three sources.

 Now we look at the relation of fringe visibility with the degree of path indistinguishability . For that we ignore the overall scaling and write the total positive component of electric field at the point of detection as \begin{align}
     \hat{E}^{(+)} = \hat{a}_1e^{i\phi_1} + \hat{a}_2e^{i\phi_2} + \hat{a}_3e^{i\phi_3}
 \end{align}
Here the phases $\phi_1$, $\phi_2$ and $\phi_3$ are acquired during the propagation of the field from source to the point of detection. So, the probability of the photon being detected at this point is
\begin{align}
   & \text{Tr}(\hat{E}^{(-)} \hat{E}^{(+)}\hat{\rho} ) = \text{Tr}\left[\left(\hat{a}^{\dagger}_1e^{-i\phi_1} + \hat{a}^{\dagger}_2e^{-i\phi_2} + \hat{a}^{\dagger}_3e^{-i\phi_3}\right)\left(\hat{a}_1e^{i\phi_1} + \hat{a}_2e^{i\phi_2} + \hat{a}_3e^{i\phi_3}\right)\hat{\rho}\right] \nonumber \\  &= \rho_{11} + \rho_{22} + \rho_{33} + 2\lbrace|\rho_{21}|~\text{cos}(\phi_{21}) + |\rho_{31}|~\text{cos}(\phi_{31}) + |\rho_{32}|~\text{cos}(\phi_{32})\rbrace \label{vis}
\end{align}
 where $\phi_{ij}=\phi_i-\phi_j$. The visibility of interference fringe is defined as\cite{Notation1} \begin{align}
     \mathcal{V}=\frac{I_{\text{max}}-I_{\text{min}}}{I_{\text{max}}+I_{\text{min}}} \label{defvis}
 \end{align}
 In Eq.\ref{vis} the angle difference $\phi_{ij}$ can be varied to get $I_{\text{max}}$ and $I_{\text{min}}$. Then from Eq.\ref{g12}, Eq.\ref{g13}, Eq.\ref{g23} and Eq.\ref{defvis} we get \begin{align}
     \mathcal{V}&=\frac{2\left(|\rho_{21}|+|\rho_{31}|+|\rho_{32}|\right)}{\rho_{11}+\rho_{22}+\rho_{33}}\nonumber\\ &= 2\left(|g^{(1)}(x_1,x_2)|~(\rho_{11}\rho_{22})^{\frac{1}{2}}+|g^{(1)}(x_1,x_3)|~(\rho_{11}\rho_{33})^{\frac{1}{2}}+|g^{(1)}(x_2,x_3)|~(\rho_{22}\rho_{33})^{\frac{1}{2}}\right) \label{nice}
 \end{align}
Using the inequality \begin{equation} 2\sqrt{I_1I_2} ~\leq ~I_1 + I_2 \label{id}\end{equation} we can write \begin{equation} 2\sqrt{\rho_{11}\rho_{22}} \leq \rho_{11} + \rho_{22} \label{hf}\end{equation} as $\rho_{ii}>0$ for i=1,2,3 we can write Eq.\ref{hf} as \begin{align} 2\sqrt{\rho_{11}\rho_{22}}~ &\leq~ \rho_{11} + \rho_{22} + \rho_{33}~ =~1 \nonumber \\ \sqrt{\rho_{11}\rho_{22}}~&\leq~\frac{1}{2} \label{2}\end{align} 
 Similarly, \begin{equation}
     \sqrt{\rho_{11}\rho_{33}}~\leq~\frac{1}{2}~~~\text{and}~~~\sqrt{\rho_{22}\rho_{33}}~\leq~\frac{1}{2}\label{3}
 \end{equation}
 Using Eq.\ref{2} and Eq.\ref{3} in Eq.\ref{nice} we get 
 \begin{align}
      \mathcal{V} ~&\leq ~\left(|g^{(1)}(x_1,x_2)|+|g^{(1)}(x_1,x_3)|+|g^{(1)}(x_2,x_3)|\right)\label{3bn}\\
      &\leq~\left(3~P_{ID}\right)
 \end{align}
Here we see  that the fringe visibility is related to the sum of modulus of two point coherence functions for all three possible pairs of slits which is reminiscent of the famous Born's rule of multi-source interference. Very simple calculation shows how according to Born's rule in quantum mechanics, the multi-slit interference experiment's fringe intensity is nothing but the sum of contribution's of all possible pairs of slits\cite{Triple}. In that paper\cite{Triple} they experimentally  proved the validity of Born's rule for three slit experiments that rules out the possibility of any multi-path, or higher order interference.

\section{Coherence and indistinguishability for interference with N sources}
Now we generalise our result for N sources where we are interested in detection of one photon. For the generalisation of the problem we use a new notation for simplification of the calculation. We denote the state $|1\rangle_1\otimes|0\rangle_2\otimes|0\rangle_3\otimes|0\rangle_4.. |0\rangle_N $ simply as $||1\rangle$. So when the photon will be generated by the m th source the state will be denoted as $||m\rangle$ which in our old notation would be $|0\rangle_1\otimes|0\rangle_2\otimes|0\rangle_3\otimes...|1\rangle_m\otimes... |0\rangle_N $ and so on. So the state of the photon for this case can be written as \begin{equation}
   |\psi\rangle = \sum\limits_{i=1}^{N}\alpha_i||i\rangle
\end{equation}
 where $|\alpha_i|^2$ is the probability of the state $\psi$ being in the i th state. Now for this the density matrix can be written as \begin{equation}
      \hat{\rho}_{ID} = |\psi\rangle\langle \psi|=\sum\limits_{i,j=1}^N\alpha_i\alpha_j^*||i\rangle\langle j||
 \end{equation}
 and likewise we denote the diagonal form of the density as \begin{equation}
      \hat{\rho}_{D}=\sum\limits_{i=1}^N\alpha_i\alpha_i^*||i\rangle\langle i|| =\sum\limits_{i=1}^N|\alpha_i|^2||i\rangle\langle i||
 \end{equation}
 Now any general density matrix can be written as \begin{align}
      \hat{\rho} &= \sum\limits_{i,j=1}^N\rho_{ij}||i\rangle\langle j|| \label{rho}
 \end{align}
 Now by decomposing this density matrix in terms of the $\rho_D$ and $\rho_{ID}$ we can write \begin{align}
      \hat{\rho} &= P_{ID} \hat{\rho}_{ID}+P_D \hat{\rho}_D \nonumber\\
     &=P_{ID}\sum\limits_{i,j=1}^N\alpha_i\alpha_j^*||i\rangle\langle j||+P_D\sum\limits_{i=1}^N|\alpha_i|^2||i\rangle\langle i|| \\ &= \sum\limits_{i=1}^N|\alpha_i|^2||i\rangle\langle i|| + P_{ID}\sum\limits_{i\neq j=1;}^N\alpha_i\alpha_j^*||i\rangle\langle j|| \label{rhogen}
 \end{align}
 Now by comparing Eq.\ref{rhogen} and Eq.\ref{rho} we can write \begin{equation}
     \rho_{ii}=|\alpha_i|^2~~~\text{and}~~~\rho_{ij}=P_{ID}\alpha_i\alpha_j^* \label{1}
 \end{equation}
 By hermiticity of density matrix we can write \begin{equation}
     \rho_{ji}=\rho_{ij}^*=P_{ID}\alpha_i^*\alpha_j \label{222}
 \end{equation}
 So, From Eq.\ref{1} and Eq.\ref{222} we can write,
 \begin{align}
\rho_{ij}\rho_{ji}=P_{ID}^2\rho_{ii}\rho_{jj} \label{333}
 \end{align}
 From Eq.\ref{1} and \ref{333} we write
 \begin{align}
     (\alpha_i\alpha_j^*)^2 &=\frac{\rho_{ij}^2}{P_{ID}^2} \nonumber\\ &=\frac{\rho_{ij}^2\times\rho_{ii}\rho_{jj}}{\rho_{ij}\rho_{ji}} \label{4}
 \end{align}
 Due to hermiticity of density matrix \begin{equation}
     \rho_{ij}=\rho_{ji}e^{2i\cdot \text{arg}(\rho_{ij})}\label{5}
 \end{equation}
 Using Eq.\ref{4}, Eq.\ref{5} and Eq.\ref{1} we can get,
\begin{align}
     P_{ID}=\frac{\rho_{ij}}{\sqrt{\rho_{ii}\rho_{jj}}}e^{-i\cdot\text{arg}(\rho_{ij})}=\frac{|\rho_{ij}|}{\sqrt{\rho_{ii}\rho_{jj}}} =\frac{|\rho_{ji}|}{\sqrt{\rho_{ii}\rho_{jj}}}\label{pidn}
 \end{align}
 $P_{ID}$ can be expressed in $^NC_2$ equivalent ways considering all possible pairs of N sources. This result boils down to Eq.\ref{3pidd} for N=3 case and L. Mandel's result\cite{Mandel} for N=2 case. Eq.\ref{pidn} suggests that the degree of path indistinguishability are equal for all possible $^NC_2$ pairs of sources.

 Now to calculate the degree of coherence we denote the positive frequency part of the single-mode electric fields of these sources as \begin{equation}
     \hat{ E}^{(+)}(r_j)=K \hat{a}_j~~~~\text{where}~~j=1,2,...N
 \end{equation}
 Now as we have N sources we will have $^NC_2$ pairs of sources for which we can calculate the second order coherence function. We calculate the general second order coherence function for a pair of points  $(x_i,x_j)$ from Eq.\ref{gdef} and Eq.\ref{rho} as
 \begin{align}
\nonumber   G^{(1)}(x_i,x_j)&=|K|^2 \text{Tr}
( \hat{a}_i^{\dagger} \hat{a}_j \hat{\rho})= |K|^2\sum\limits_{n=1}^N\langle n|| \hat{a}_i^{\dagger} \hat{a}_j \hat{\rho} ||n\rangle\\\nonumber &= |K|^2\sum\limits_{n=1}^N\langle n|| \hat{a}_i^{\dagger} \hat{a}_j\left(\sum\limits_{l,k=1}^{N}\rho_{lk}||l\rangle\langle k|| \right)||n\rangle\\ &= |K|^2\sum\limits_{k,l=1}^{N}\rho_{lk}\delta_{ki}\delta_{jl}~= |K|^2\rho_{ji}\label{66666}\end{align}
From Eq.\ref{66666} and Eq.\ref{gdef} we see that \begin{equation}
     g^{(1)}(x_i,x_j)=\frac{\rho_{ji}}{\sqrt{\rho_{ii}\rho_{jj}}}. \label{6}\end{equation}
 We also note any higher order coherence function than this will be zero for this generalised case also.
From Eq.\ref{pidn} and Eq.\ref{6} one sees that these pairwise second order coherence functions are equal to the modulus of degree of indistinguishability,
 \begin{equation}
   |g^{(1)}(x_i,x_j)|=P_{ID}  \label{finr}
 \end{equation}

 Therefore, for N single-mode fields when we consider the detection of a single photon only we see that the degree of coherence is exactly equal to the modulus of  degree of indistinguishability when we consider all possible pairs of the sources. Now to relate degree of indistinguishability with the degree of coherence we denote the total positive component at the point of detection without the scaling factor as \begin{equation}
     \hat{E}^{(+)}~=~\sum\limits_{m=1}^{N}\hat{a}_me^{i\phi_m}
 \end{equation}
 where $\phi_n$ is the phase acquired by the field while propagating to the point of detection from the n th source. We denote the intensity as well as the detection probability at this point of detection as expectation of the operator $\hat{E}^{(-)}\hat{E}^{(+)}$ as \begin{align}
   \label{98}  \text{Tr}\left(\hat{E}^{(-)}\hat{E}^{(+)}\rho\right) =  \sum\limits_{i=1}^N\rho_{ii}~+2\sum\limits_{i,j=1;i>j}^N|\rho_{ij}|~\text{cos}(\phi_{ij})
 \end{align}
 From Eq.\ref{defvis}, Eq.\ref{id} and Eq.\ref{98} we write the visibility as \begin{align}
     \mathcal{V}~&=~2~\frac{\sum\limits_{i,j=1;i>j}^N|\rho_{ij}|}{\sum\limits_{i=1}^N|\rho_{ii}|}  
     ~=~2\sum\limits_{i,j=1;i>j}^N|g^{(1)}(x_i,x_j)|(\rho_{ii}\rho_{jj})^{\frac{1}{2}}~\text{where}~i>j \label{nform}\\ &\leq \sum\limits_{i,j=1;i>j}^N(|g^{(1)}(x_i,x_j)|) \label{ngen1}\\ &\leq~ ^NC_2P_{ID}\label{ngen2}
 \end{align}
 Eq.\ref{nform} is the general form for interference fringe visibility of  N single-mode fields. Eq.\ref{ngen1} and Eq.\ref{ngen2} are the generalised relations between the fringe visibility , degree of coherence and degree of indistinguishability. Here also as expected from the results of three fields we get relation between visibility and degree of coherence that resembles the Born's rule.\raggedbottom
 \section{Conclusion}
 In this article we present a generalised relations between degree of indistinguishability and the degree of coherence. We also see how these quantities are related to the visibility of interference fringe for N single-mode fields for detection of one photon only. We also note the consistency of these results with  Born's rule for the interference fringe . These results present a good picture or relations between wave and particle nature of photon.

\end{document}